\documentclass[floats,floatfix,amssymb,prd,twocolumn,superscriptaddress,nofootinbib]{revtex4-2}
	
\makeatletter
\def\l@subsubsection#1#2{}
\def\l@subsubsubsection#1#2{}
\makeatother

\usepackage{graphicx,amssymb,amsmath,amsthm,amsfonts,epsfig,epsf,fixmath}
\usepackage[usenames]{color}
\usepackage{epstopdf}
\usepackage[normalem]{ulem}
\usepackage{comment}
\usepackage{tabularx}
\usepackage{bm}
\usepackage{dcolumn}
\usepackage{latexsym}
\usepackage{rotating}
\usepackage{longtable}

\usepackage{enumerate}
\usepackage{tensor,multirow}
\usepackage{url}
\usepackage[dvipsnames]{xcolor}
\usepackage[unicode]{hyperref}
\usepackage[caption=false]{subfig}
\usepackage{float}
\usepackage{lipsum}  

\hypersetup{colorlinks=true, citecolor=MidnightBlue,
            linkcolor=MidnightBlue, urlcolor=MidnightBlue, linktocpage=true}

\newcommand{\tn}{\textnormal}

\newcommand{\GSSI}{Gran Sasso Science Institute (GSSI), I-67100 L’Aquila, Italy}
\newcommand{\GranSasso}{INFN, Laboratori Nazionali del Gran Sasso, I-67100 Assergi, Italy}

\newcommand{\milan}{\affiliation{Dipartimento di Fisica ``G. Occhialini'', 
Universit\'a degli Studi di Milano-Bicocca, Piazza della Scienza 3, 20126 Milano, Italy}}
\newcommand{\infnmilan}{\affiliation{INFN, Sezione di Milano-Bicocca, 
Piazza della Scienza 3, 20126 Milano, Italy}}

\begin{document}

\title{Bayesian parameter estimation on boson-star binary signals with a coherent inspiral template and spin-dependent quadrupolar corrections}

\author{Massimo Vaglio}
\affiliation{Dipartimento di Fisica, ``Sapienza'' Universit\`a di Roma, Piazzale
Aldo Moro 5, 00185, Roma, Italy}
\address{INFN, Sezione di Roma, Piazzale Aldo Moro 2, 00185, Roma, Italy}
 \author{Costantino Pacilio}
\milan
\infnmilan
 \author{Andrea Maselli}
\address{\GSSI}
\address{\GranSasso}
\author{Paolo Pani}
\affiliation{Dipartimento di Fisica, ``Sapienza'' Universit\`a di Roma, Piazzale
Aldo Moro 5, 00185, Roma, Italy}
\address{INFN, Sezione di Roma, Piazzale Aldo Moro 2, 00185, Roma, Italy}

\date{\today}
\begin{abstract}
Compact boson star binaries are hypothetical sources for ground-based and space gravitational-wave detectors. 
Their signal would be a messenger for novel fundamental fields and could shed light on the dark matter.
In this work, we further develop our analysis in Phys.~Rev.~D 102,~083002~(2020), aimed at constraining the properties of these objects with future observations. We use a coherent waveform template for the inspiral stage of boson star binaries with large quartic self interactions, including tidal deformability and the nonlinear dependence of the quadrupole moments on the spin in terms of the fundamental couplings of the scalar field theory. Performing a Bayesian analysis, we investigate the ability of a third-generation gravitational-wave detector such as the Einstein Telescope to distinguish these exotic sources from black holes and infer constraints on the fundamental couplings of the model.
\end{abstract}

\maketitle

\section{Introduction}
Boson stars~(BSs) are self-gravitating condensates of a bosonic field~\cite{Kaup:1968zz,Ruffini:1969qy,Colpi:1986ye} (see Refs.~\cite{Jetzer:1991jr,Schunck:2003kk,Liebling:2012fv} for some reviews). Their properties such as maximum mass, radius, compactness, spin- and tidally-induced deformabilities strongly depend on the self-interactions of the underlying field theory~\cite{Cardoso:2017cfl,Sennett:2017etc,Boskovic:2021nfs}.
Depending on the latter, BSs can exist in any mass range and can have a compactness comparable to or larger than that of a neutron star.
Like standard compact objects such as neutron stars and black holes~(BHs), BSs can be produced dynamically through gravitational cooling~\cite{Seidel:1993zk,DiGiovanni:2018bvo} and can form compact binaries whose coalescence (see, e.g.,~\cite{Liebling:2012fv,Palenzuela:2007dm,Palenzuela:2017kcg,Bezares:2017mzk,Sanchis-Gual:2018oui,Bezares:2022obu,Siemonsen:2023hko}) is a primary target of current and future gravitational-wave~(GW) detectors~\cite{Cardoso:2019rvt,Maggiore:2019uih,LISA:2022kgy,Barausse:2020rsu}.
In this context, it is intriguing that some of the detected GW events (e.g., GW190814~\cite{LIGOScientific:2020zkf} and GW190521~\cite{LIGOScientific:2020iuh}) might not fit naturally within the standard astrophysical formation scenarios and are compatible with more exotic origins, such as with the coalescence of two Proca stars~\cite{CalderonBustillo:2020fyi}, which are the vector counterparts~\cite{Brito:2015pxa} of the scalar BSs studied in this work.

Here we continue our investigation of the detectability of BS binaries with large quartic interactions~\cite{Colpi:1986ye} (sometimes called massive BSs), extending our previous studies~\cite{Pacilio:2020jza,Vaglio:2022flq} in various directions.
We adopt the coherent waveform template for the inspiral phase developed in Ref.~\cite{Pacilio:2020jza}, which is based on a TaylorF2 post-Newtonian~(pN) model~\cite{Damour:2000zb}, but coherently including the spin-induced quadrupole moment~\cite{poisson_gravitational_1998,PoissonWill} and the tidal deformability parameters~\cite{Flanagan:2007ix,Hinderer:2007mb,PoissonWill} as predicted by the scalar field theory with quartic interactions. Therefore, the waveform model depends only on the binary masses and spins, on the scalar-field mass, and on the quartic self-interaction. In the strongly-interacting limit, the latter two combine into a single parameter ~\cite{Pacilio:2020jza,Vaglio:2022flq}, so the waveform model contains only a single extra parameter with respect to a binary BH inspiral.
Furthermore, we take advantage of our recent results for the spin-induced quadrupole moment of massive BSs~\cite{Vaglio:2022flq}, and include the complete nonlinear spin dependence of the 2pN phase terms in the waveform which --~together with the fact that BSs have nonzero tidal deformability~\cite{Cardoso:2017cfl,Sennett:2017etc}~-- is a key discriminator with respect to the BH case.
Finally, we implement this updated waveform model into a Bayesian parameter estimation\footnote{A similar Bayesian inference has been done in Ref.~\cite{Johnson-Mcdaniel:2018cdu} using a simple polytropic perfect-fluid model to approximate a boson star and neglecting spin effects.} using the projected sensitivity of future third-generation GW interferometers such as the Einstein Telescope~(ET)~\cite{Hild:2010id,Maggiore:2019uih}. We show that, for typical coalescence events, ET can measure the coupling constant of the scalar-field theory with percent accuracy.

Henceforth we use $G=c=1$ units.

\section{Finite size effects for Boson Star binaries}
\label{sec: lagrangian}

Stationary axisymmetric BSs are solutions
to the Einstein–Klein-Gordon equations for a complex,
massive, scalar field, minimally coupled to
the gravitational sector~\cite{Kaup:1968zz,Ruffini:1969qy,Colpi:1986ye}. The theory is described by the action 
\begin{equation*}
{\cal S}=\int d^4 x\sqrt{-g}
\left[\frac{R}{16\pi}+{\cal L}_\phi\right]\ ,
\end{equation*} 
where $R$ is the Ricci scalar, and the Lagrangian ${\cal L}_\phi$ 
governs the dynamics of the field $\phi$. Here 
we consider a family of BSs characterized quartic interactions~\cite{Colpi:1986ye} (massive BSs):
\begin{equation}
{\cal L}_\phi=-\frac{1}{2}g^{\mu\nu}\phi^*_{,\mu}\phi_{,\nu}
-\frac{1}{2}\mu^2|\phi|^2-\frac{1}{4}\sigma|\phi|^4\ .
\end{equation}
The mass parameter $\mu$ is related to the physical mass of the bosonic particle by $m_{S}=\hbar \mu$.
We shall focus on the strong coupling limit, $\sigma\gg\mu^2$, which allows for very compact equilibrium configurations.
In this limit the properties of an equilibrium configuration depend on the coupling constants $\mu$ and $\sigma$ only through their
combination $M_B=\sqrt{\sigma}/\mu^2$, which has dimensions of a mass~\cite{Pacilio:2020jza,Vaglio:2022flq}.
Moreover, we introduce the dimensionless 
ratio $\beta~\equiv M/M_B$ (with $M$ being the BS mass), which is convenient to parameterize
the properties of the stellar structure~\cite{Vaglio:2022flq,Ryan:1996nk}.

Finite-size effects related to the BS structure differ significantly 
from those of BHs, and introduce distinctive features within 
the GW signals emitted by coalescing systems, which can 
be exploited to disentangle the two families of compact objects. 
Such signatures play different roles during the inspiral stages 
of a binary. Far from the merger, at low frequencies, sources 
behave as point 
particles and finite size effects are effaced~\cite{blanchet_gravitational_2014}.  
As the binary approaches the last phase of the coalescence, 
finite-size effects, induced by stellar structure, become more 
relevant and are strong enough to modify the system 
orbital evolution and its GW emission~\cite{poisson_gravitational_1998,hinderer_tidal_2010, Favata:2013rwa}. These 
modifications carry precious information on the nature of 
the binary components, and allow studying a variety of 
fundamental physics problems, from the properties of 
dense nuclear matter~\cite{maselli_constraining_2013,agathos_constraining_2015,chatziioannou_neutron_2020}, to the nature of BHs~\cite{Cardoso:2019rvt,Cardoso:2017cfl, Barack:2018yly} and their environments~\cite{Macedo:2013qea,Barausse:2014tra,Cardoso:2019rou,Cardoso:2019upw,DeLuca:2021ite,DeLuca:2022xlz}.

Here, adopting a pN approximation of the inspiral waveform including finite-size effects, we focus on two smoking gun signatures 
given by: (i)~nonvanishing tidal deformabilities~\cite{Cardoso:2017cfl,Sennett:2017etc}
and (ii) spin induced multipole moments~\cite{Ryan:1996nk,Pacilio:2020jza}
which differ from their Kerr counterparts.  
We model finite-size effects by additional contributions to the phase of the waveform, 
$\varphi=\varphi_{\rm pp}+\varphi_\tn{T}+\varphi_{\kappa}$, 
where $\varphi=\varphi_{\rm pp}$ is the point-particle phase and
$\varphi_\tn{T}$ and $\varphi_{\kappa}$ identify 
the phase contributions of tidal interactions and of 
the spin-induced multipole moments, respectively.

As in the case of neutron-star binaries, deformations induced on BSs 
by the external tidal field of a companion leave a footprint 
on the emitted waveform encoded by the dimensionless 
tidal Love numbers $\lambda_\ell$, which depend on the coupling constant of the theory and on the object's compactness~\cite{Cardoso:2017cfl,Sennett:2017etc}.
The leading contribution to GW emission is given by the 
quadrupolar coefficient $\lambda_2$~\cite{Hinderer:2007mb}, which determines the dimensionless tidal deformability 
\begin{equation}
\Lambda=\frac{2}{3}\lambda_2 \frac{R^{5}}{M^5}=\frac{2}{3}\lambda_2 
{\cal C}^{-5}\ ,
\end{equation}
where $M$, $R$ and ${\cal C}=M/R$ are the BS mass, radius, and compactness. 
The phase contribution $\varphi_\tn{T}$ depends on the 
tidal deformabilities $\Lambda_{1,2}$ of both binary components, 
and at the leading order enters the GW phase at 5pN with a coefficient~\cite{Favata:2013rwa} (see Eq.~\eqref{eq:PN_expansion} below and Appendix~\ref{sec:pnappendix})
\begin{align}
\varphi_{\tn{T},5}
=-12[(1+7\eta-&31\eta^2)(\Lambda_1+\Lambda_2)\nonumber\\
&+\delta(1+9\eta-11\eta^2)(\Lambda_1-\Lambda_2)]\ ,
\end{align}
where 
$\eta=m_1 m_2/(m_1+m_2)^2$ is 
the symmetric mass ratio, $\delta=(m_1-m_2)/m$, and $m=m_1+m_2$ is the  total mass of the binary.

A second key difference between BHs and BSs lies in their  
spin-induced multipolar structure.
In General Relativity, stationary BHs are uniquely 
described by the Kerr solution, whose multipolar structure is  completely determined by the mass and spin through the compact relations~\cite{hansen_multipole_1974}
\begin{align}
Q_\ell^{BH}+iS_\ell^{BH}=M^{\ell+1}(i\chi)^\ell\ ,\label{BHmoments}
\end{align}
where $Q_\ell$ and $S_\ell$ are two sets of mass and 
current moments, respectively, and $\chi= J/M^2$ is the BH dimensionless 
spin parameter, being $J$ its angular momentum~\cite{hansen_multipole_1974}. 
All other multipole moments vanish identically, as dictated by the axisymmetry and equatorial symmetry of the Kerr metric.
For a rotating BSs, the multipole moments depend on the mass and spin but also on the scalar couplings. For example, the quadrupole moment (which is the dominant parameter, entering at 2pN order in the GW phase), can be expressed as 
\begin{align}
Q_2=-\kappa_2M^3\chi^2\ ,
\end{align}
where $\kappa_2=\kappa_2(\chi,\beta)$. The 
dependence of $\kappa_2$ on the BS mass and spin, as well on the coupling parameter $M_B$ has been investigated in details in Ref.~\cite{Vaglio:2022flq}.
In particular, it is found that $\kappa_2$ decreases 
as the BS compactness increases and, for slowly 
spinning configurations close to the maximum 
compactness, has a lower cutoff $\kappa_2\approx20$,  
to be contrasted with the BH case $\kappa_2^\tn{BH}=1$.
Hereafter we ignore 
multipoles with $\ell>2$ in the signal, focusing on the leading  
contribution given by the quadrupole coefficient $\kappa_2$.  
The latter introduces a correction to the point-particle waveform phase which enters at 2pN order together with 
quadratic terms in the BS spin parameter~\cite{poisson_gravitational_1998, barack_using_2007}, with a leading-order coefficient proportional to (see Eq.~\eqref{eq:PN_expansion} below and Appendix~\ref{sec:pnappendix})
\begin{equation}
\varphi_{\kappa,2} = - \frac{50}{m^2}(\kappa_1m_1^2\chi_1^2+\kappa_2m_2^2\chi_2^2)\ .
\end{equation}

\section{Binary waveform model}
\label{sec:taylorbs}
Our waveform model builds on the frequency-domain 
TaylorF2 template~\cite{Damour:2000zb}, which describes the GW inspiral signal 
of a binary system on quasicircular orbits,
\begin{equation}
\tilde{h}(f,\bm{\theta})=\mathcal{A}(f,\bm{\theta})e^{-i\varphi(f,\bm{\theta})}\ ,
\end{equation}
where $f$ is the frequency and $\bm{\theta}$ identifies the set of source parameters.  
Both the amplitude and the phase of the signal are expanded 
as a power series in the parameter $x=(\pi f m)^{2/3}$. A term proportional to $x^n$ 
corresponds to the $n$-pN order of the approximation. 
Here we retain only the 0pN (Newtonian) term for the amplitude 
\begin{equation}
{\cal A}={\cal C}\sqrt{\frac{5\pi}{96}}\frac{{\cal M}^{5/6}}{\pi^{2/3}d_\tn{L}}f^{-7/6}\ ,    
\end{equation}
where $\mathcal{M}=(m_1m_2)^{3/5}/m^{1/5}$ is the 
chirp mass, and $d_\tn{L}$ the binary luminosity distance.
The geometric factor 
\begin{equation}   
{\cal C}=\sqrt{(1+\cos^2\iota)^2F_+^2+4\cos^2\iota F_\times^2}\ ,
\end{equation}
depends on the antenna pattern functions 
$F_{+,\times}\equiv F_{+,\times}(ra,dec,\psi)$ and 
encodes the information about the right ascension 
$ra$ and declination $dec$ of the source, the polarization angle $\psi$, and the inclination angle $\iota$ between the binary orbital 
angular momentum and the line of sight of the source~\cite{schutz_antenna_1987}. 
We expand the phase including contributions up to 6pN order:
\begin{align}
&\varphi=2\pi ft_c-\varphi_c-\pi/4 +\frac{3}{128 \eta x^{5/2}} \times \nonumber\\
&  \left( \sum_{i=0}^{7} \varphi_{{\rm pp},i/2}x^{i/2} +\varphi_{\kappa,2}x^2+\varphi_{\kappa,3}x^3 +\varphi_\tn{T,5}x^5+\varphi_\tn{T,6}x^6 \right) \ ,
\label{eq:PN_expansion}
\end{align}
where $t_c$ and $\varphi_c$ are the time and phase at the 
coalescence, respectively. 
Specifically, we consider: (i)~point particle coefficients $\varphi_{{\rm pp},i}$ 
including spin-orbit and spin-spin terms up to 
3.5pN order~\cite{Khan:2015jqa}, (ii) 
2pN and 3pN spin-induced quadrupole corrections 
$\varphi_{\kappa,2}$ and $\varphi_{\kappa,3}$ as computed in 
\cite{Krishnendu:2017shb}, (iii)~tidal contributions 
$\varphi_\tn{T,5}$ and $\varphi_\tn{T,6}$ entering at 5pN and 
6pN order~\cite{Lackey:2014fwa}.  
We refer the reader to Appendix~\ref{sec:pnappendix} for their 
explicit form.

Ref.~\cite{Sennett:2017etc} found a simple
semi-analytic expression relating $\beta=M/M_B$ to 
the static tidal deformability $\Lambda$, which can
be inverted in the form
\begin{equation}
\log_{10}\Lambda= \sum_{k=0}^4 \alpha_k \beta^k\ ,\label{tidal}
\end{equation}
where 
\begin{align}
&\alpha_0 = +7.5\ , &&\alpha_1 = -9.0\ , \nonumber\\
&\alpha_2 = -4.4\times 10^3\ , &&\alpha_3 = +1.0\times10^5\ , \nonumber\\
&\alpha_4 = -7.7\times10^5\ .
\end{align}

We will use this relation also for binary systems with spinning BSs, i.e. neglecting the possible (subleading) dependence 
of $\Lambda$ on the BS spin (see Ref.~\cite{Castro:2022mpw} for an estimate of the relevance of these terms in the neutron-star case).

The BS quadrupole moment can also be determined in terms of 
$\beta$ and of the dimensionless spin parameter $\chi$, 
$\kappa_2=\kappa_2 \left(\beta,\chi\right)$, through a cubic bi-spline 
interpolation of the data derived in Ref.~\cite{Vaglio:2022flq} 
on a two-dimensional $(\beta, \chi)$ grid with $9 \times 30$ points in the range $0.02\leq\beta\leq0.06$ and $0\leq\chi\leq 3$. 
The errors in the 
fit~\eqref{tidal} and on the numerical interpolation of $\kappa_2$ for the tidal and 
quadrupole terms are discussed in Appendix~\ref{sec:fitappendix} 
and are below $10\%$ and $5\%$, respectively, 
within the entire grid domain. 

The waveform so constructed can be fully specified by 9 intrinsic parameters and 7 extrinsic parameters. 
The intrinsic parameters are the chirp mass 
${\cal M}$, the symmetric mass ratio $\eta$, 
the spin magnitudes $a_{1,2}$, the tilt angles between 
the spin directions and the orbital angular momentum $\theta_{1,2}$, 
the angle $\delta \phi$ between the two spin vectors, 
the angle $\theta_{JL}$
between the orbital and total angular momentum of the binary and
the BS effective coupling $M_B=\sqrt{\sigma}/\mu^2$. 
The latter sets the mass 
scale for the two compact objects, and determines their tidal 
deformabilities and the spin-induced quadrupole moments~\cite{Pacilio:2020jza,Vaglio:2022flq}. 
The extrinsic parameters are the sky localization angles 
$(ra,dec)$, the luminosity distance $d_\tn{L}$, the 
inclination and the polarization angles $(\iota,\psi)$, as well 
as the time and phase at the coalescence $t_c$ and $\varphi_c$. 

Our template is publicly available at~\cite{webpage,webpageA}.

\section{Parameter estimation settings}
\label{sec:params}
We simulate injection and recovery of the waveform parameters
with the open software \texttt{bilby}~\cite{Ashton:2018jfp,Romero-Shaw:2020owr}. We project the signals into a triangular ET detector with the ET-D 
sensitivity~\cite{hild_xylophone_2010,Hild:2010id} and inject the signals into white noise 
by adjusting the luminosity distance to the desired signal-to-noise ratio~(SNR).

We expect the sky-position angles $(ra,dec)$, the inclination $\iota$
and the polarization angle $\psi$ to be weakly correlated 
with the intrinsic parameters. Since our aim 
is to investigate the recovery of the BS coupling $M_B$ and how this affects the recovery of the other intrinsic parameters, for simplicity
we remove $(ra,dec,\iota,\psi)$ from the set of waveform parameters. Moreover, we focus on orbital configurations with spins (anti-)aligned with the angular momentum. Therefore, we restrict the inference to the set of parameters $\bm{\theta}=\{{\cal M},\eta,\chi_1,\chi_2,M_B,d_L,t_c,\phi_c\}$, where $\chi_i$ is the projection of the $i$-th spin along the z direction and $M_B$ is the BS characteristic scale introduced in Sec.~\ref{sec: lagrangian}.
We use as injection parameters for the simulated systems the sky localization angles $(ra,dec)$ inferred for the event GW170817~\cite{LIGOScientific:2018hze,Coulter:2017wya} and we fix the inclination and polarization angles to $(\iota,\psi)=(0.4,2.659)$. We sample the posteriors with a \texttt{dynesty} sampler using $2500$ live points and we use analytic marginalization over $(d_\tn{L},t_c, \phi_c)$.

\begin{figure}[thp]
\centering
    \includegraphics[width=0.49\textwidth]{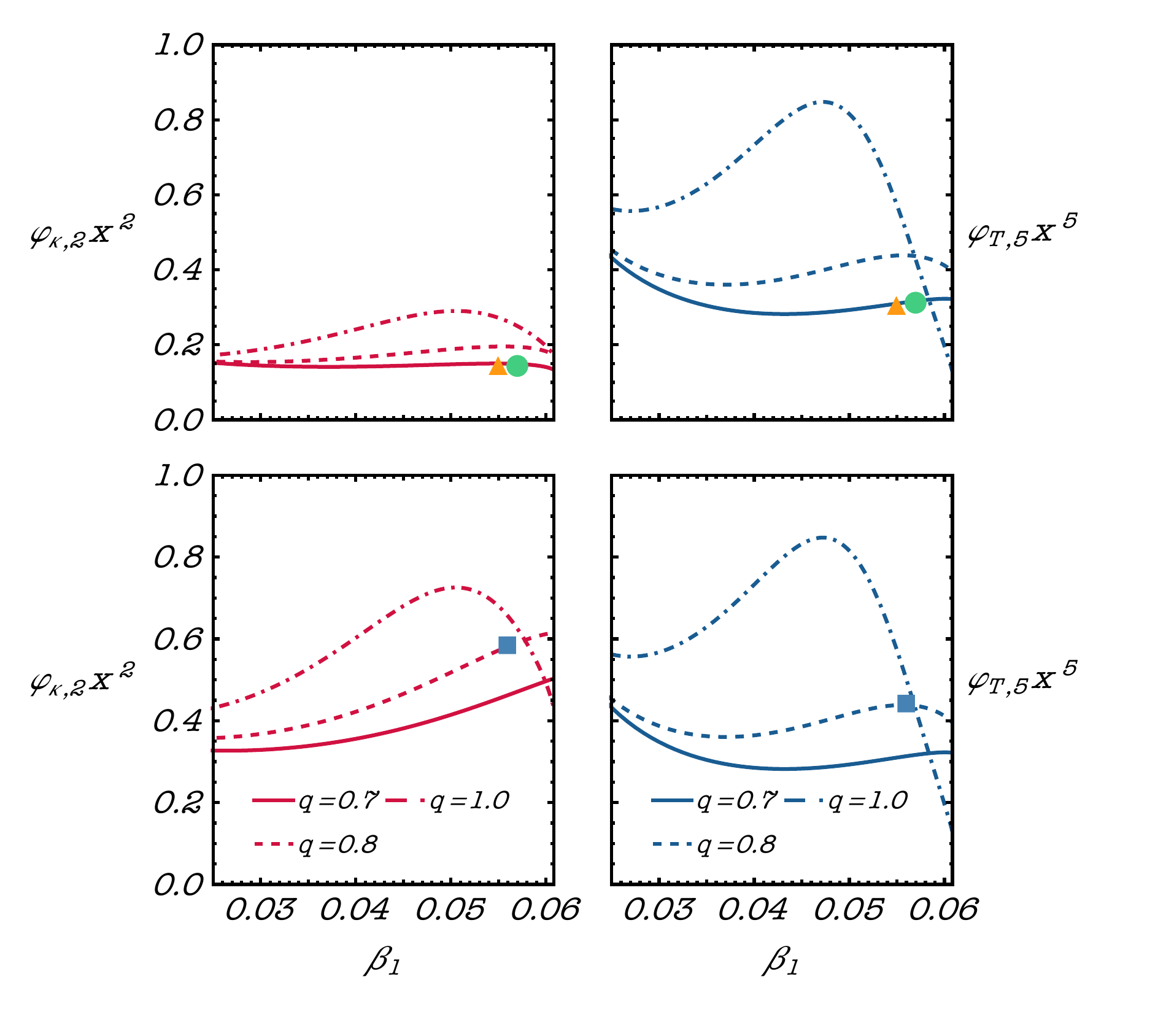}
    \caption{Phase contribution of the leading quadrupole (2pN) and  
    tidal (5pN) corrections normalized to the Newtonian term, evaluated at 
    $f=f_\text{\tiny Roche}(\beta_1,q)$ for individual spins $\chi_1=0.2,\chi_2=0.1$ (top) and $\chi_1=0.05,\chi_2=0.35$ (bottom). The relative amplitudes are shown as functions of 
    the primary rescaled mass $\beta_1=m_1/M_B$ and for different values of the mass ratio 
    $q$. Circle, square, and triangle dots represent the values for the first, second, and third system in Table~\ref{tab:parameters}, respectively.
    }
    \label{fig:pn_contribution}
\end{figure}

Assuming Gaussian noise, the likelihood can be expressed as
\begin{align}
\mathcal{L}(d\vert\bm{\theta})
=\tn{exp}\Bigl[-\frac{1}{2}(d-h(\bm{\theta})|d-h(\bm{\theta}))\Bigr]\ ,
\end{align}
where the inner product is defined by
\begin{align}
(h_1,h_2)\equiv4\, {\rm Re}\int_{f_\tn{min}}^{f_\tn{max}} \frac{\tilde{h}_1(f)\tilde{h}_2^*(f)}{S_n(f)} df\ ,\label{math:scalar}
\end{align}
with $S_n(f)$ being the on one-side power spectral density. We fix the minimum frequency to $f_{\rm min}=5$ Hz. The choice of the maximum frequency $f_{\rm max}$ requires more care and depends on the intrinsic parameters of the injected event. Indeed, similarly to neutron stars, we need to take into 
account the effect of tidal deformations 
during the inspiral. In order to be conservative we take $f_{\rm max}$ as the frequency at which the tidal force of the more compact BS overcomes the self gravity of the less compact, leading to its tidal disruption well before merger. In the Newtonian approximation, this happens when the binary 
orbital separation is comparable to the Roche radius~\cite{roche_limit_1849}
\begin{equation}
r_\text{{\tiny Roche}}\sim \gamma r_2  \left(\frac{m_1}{m_2}\right)^\frac{1}{3}\ ,\label{math:roche}
\end{equation}
where $r_2$ is the radius of the secondary BS. The coefficient $\gamma$ varies from the rigid body limit $\gamma=1.26$ to the fluid body limit $\gamma=2.44$.
\begin{figure}[thp]
\centering
    \includegraphics[width=0.5\textwidth]{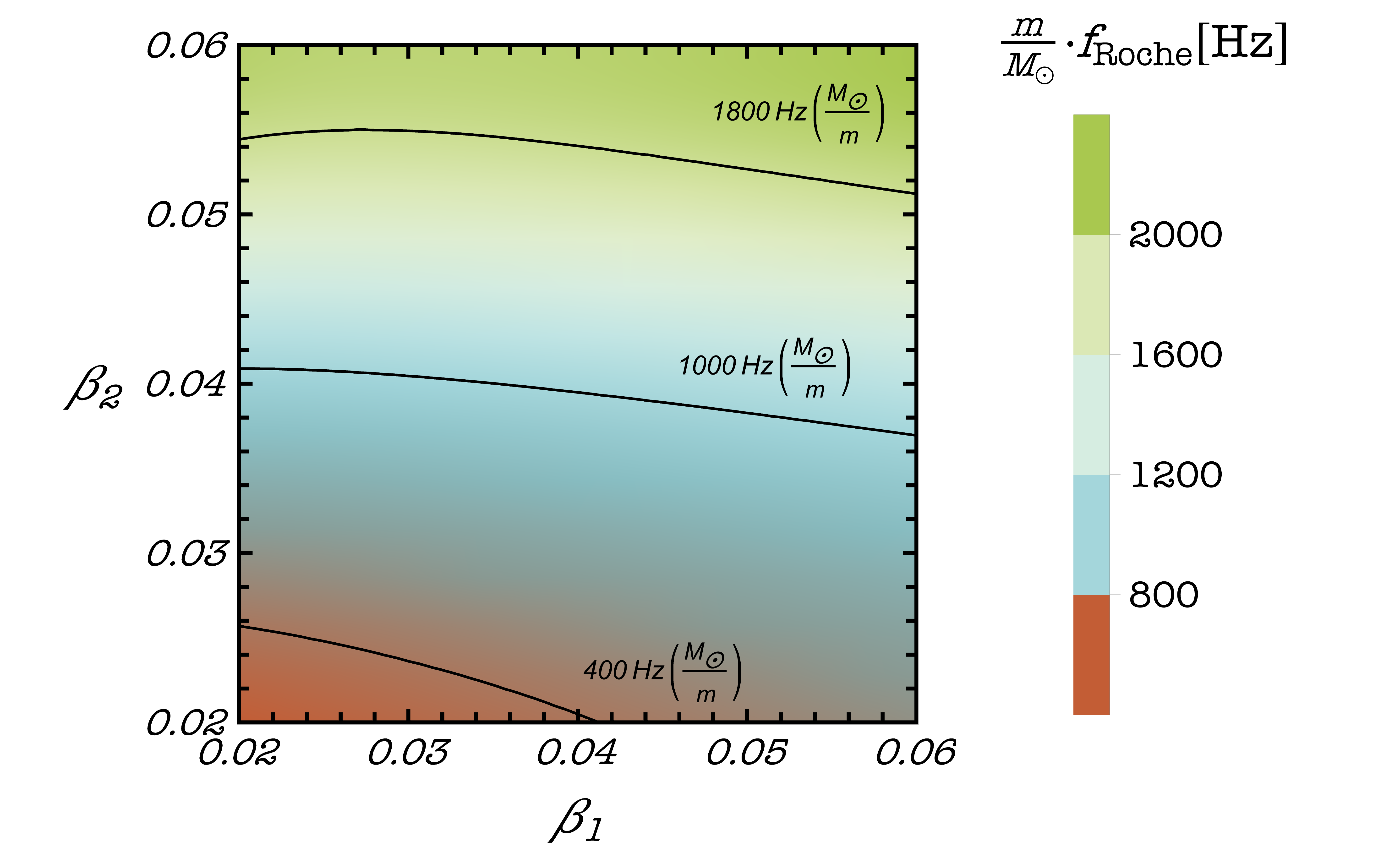}
    \caption{Density plot showing the product $m/M_\odot \cdot f_{\text{Roche}}$ as a function of the rescaled masses $\beta_1$ and $\beta_2$ of the binary components. Contour lines at three different fixed values are shown as black continuous 
    curves.}
    \label{fig:roche_density}
\end{figure}
The corresponding Roche frequency is given by
\begin{equation}
f_\tn{\rm Roche}=\frac{1}{\pi m}\sqrt{3+q+3q^{-1}+q^{-2}}\left(\frac{\mathcal{C}_2}{\gamma}\right)^\frac{3}{2}\ ,
\end{equation}
where $q=m_2/m_1\leq1$ is the mass ratio,
and $\mathcal{C}_2$ is the compactness of the less 
compact star in the binary. 
The density plot in Fig.~\ref{fig:roche_density} shows 
values of constant $m/M_\odot \cdot f_{\text{Roche}}$ as 
a function of the rescaled masses $\beta_{i=1,2}$.
Hereafter\footnote{This is a conservative assumption. 
Indeed, we have also studied the case in which $f_\tn{max}$ 
is given by the contact frequency $f_\tn{contact}$, defined so that the distance between the 
centers of the objects is equal to the sum of 
their radii. We also included pN
corrections accounting for the tidal deformation 
induced on the BS shape, encoded by the superficial 
Love numbers~\cite{Johnson-Mcdaniel:2018cdu}. 
For all binaries we considered, $f_\tn{contact}>f_\tn{\rm Roche}$, and including higher frequencies generically improves the accuracy of the inference. We also fix $\gamma=2.44$, being again conservative, since smaller values of $\gamma$ correspond to a higher cut-off frequency.}
we fix $f_\tn{max}=f_\tn{\rm Roche}$. 
As a further check of the overall consistency of our framework,
we have verified that, despite the large values of the tidal deformabilitites
$\Lambda$ and of the quadrupole moments $\kappa_2$, tidal and quadrupolar contributions to the GW phase appearing in Eq.~\eqref{eq:PN_expansion},  
are always smaller than the Newtonian (0pN) term $\phi_{{
\rm pp},0}=1$ at $f=f_\text{{\tiny max}}$, ensuring the validity of the 
pN expansion. The comparison between $\varphi_{{\rm T},5}$ and $\varphi_{{\kappa},2}$ is shown in Fig.~\ref{fig:pn_contribution} as a function 
of $\beta_1=m_1/M_B$ for different values of the mass ratio $q$.
This comparison shows that, depending on the binary parameters, the two effects can be comparable.


We impose flat priors on the chirp mass $\mathcal{M}$, the 
symmetric mass ratio $\eta$, the coupling $M_B$, and the spins $\chi_i$ 
within the ranges $\mathcal{M}\in[1,25]M_\odot$,  
$\eta \in [0,0.25]$, 
$M_B\in[10,500]M_\odot$, and $\chi_i\in[-1,1]$, respectively. 
\begin{figure}[thp]
\centering
    \includegraphics[width=0.49\textwidth]{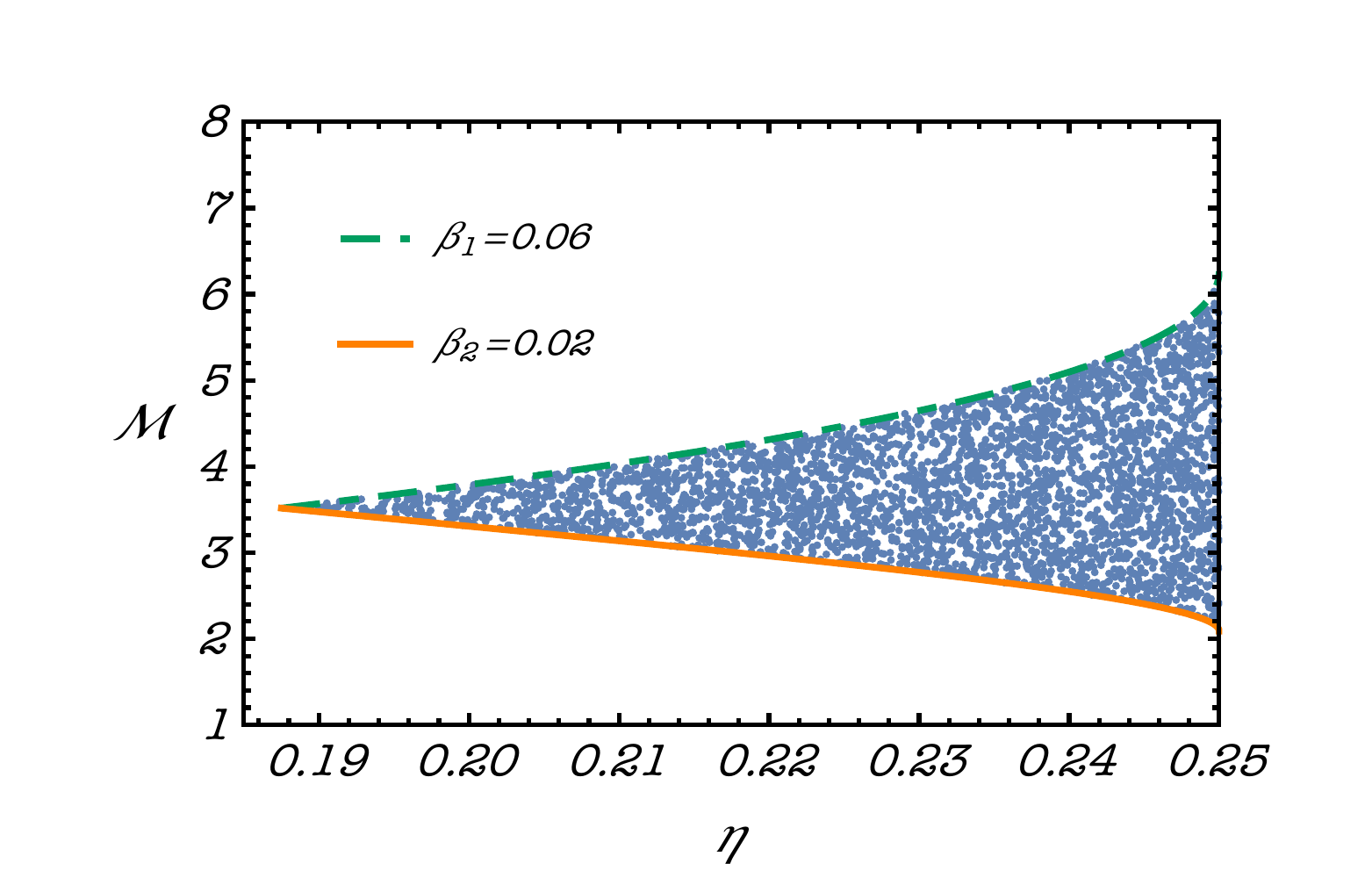}
    \caption{Uniform sampling in the 
    chirp mass $\mathcal{M}$ and in the symmetric mass 
    ratio $\eta$ with constraints on the dimensionless mass parameters 
    $0.02<\beta_2<\beta_1<0.06$ for a fixed value of $M_B=120M_\odot$. Solid and dashed 
    curves correspond to the upper and lower values of each component mass.}
    \label{fig:samples}
\end{figure}
Moreover, we require that 
the BS parameters $\beta_{i=1,2}$ satisfy the constraint 
$0.02 \leq \beta_{1,2} \leq 0.06$. 
The upper bound corresponds to the value of $\beta$ for 
which the star reaches the maximum mass allowed by 
the model for static stellar configurations 
\cite{Colpi:1986ye,Pacilio:2020jza},
while the lower edge is chosen to exclude stellar 
configurations which are too diluted. Indeed, for 
$\beta\lesssim 0.02$, the BS compactness drops 
to\footnote{Such lower value of ${\cal C}$ holds 
for static configurations, although 
slowly spinning BSs feature a similar 
bound~\cite{Ryan:1996nk}.} 
$\mathcal{C}\lesssim 0.035$, 
much less compact than a typical neutron star.
The constraints on $\beta_{1,2}$ introduce a nontrivial 
dependence on the joint prior distribution of 
${\cal M}$ and $\eta$. This is clear in 
Fig.~\ref{fig:samples} which shows 
how the uniform sampling in the 
($\mathcal{M},\eta)$ plane is cut by our prior 
assumption on $\beta$, for a representative value of $M_B$. 

\section{Results}
We show here the results of our parameter estimation 
for three representative binary BS systems with properties listed in Table~\ref{tab:parameters}, the first two having $\mathcal{M}=5M_\odot$ and the third one with $\mathcal{M}=10M_\odot$. The first and third systems share the same values of $(\eta, \chi_1, \chi_2)$, while the second system has a higher value of the mass ratio and different spin components. The value of $M_B$ for each system is chosen in such a way to have $\beta_1$ close to the upper limit $\beta_{\text{\tiny max}}=0.06$, so the primary BS is near its maximum compactness.

 To compare the results, we choose the luminosity distance $d_L$ such that ${\rm SNR}\sim 130$ for all systems. The values reported for the component masses, the chirp mass, and the coupling $M_B$ are all expressed in the detector frame\footnote{Notice that because both $m_i$ and $M_B$ are given in the detector frame, their ratio $\beta_i$ is redshift-independent, and so are dimensionless quantities such as $\Lambda(\beta)$ and $\kappa_2(\beta,\chi)$.}.
\begin{table}[thp]
\scriptsize
\begin{tabularx}{\linewidth}{
>{\hsize=2.0\hsize\linewidth=\hsize}X
>{\hsize=1.8\hsize\linewidth=\hsize}X
>{\hsize=.5\hsize\linewidth=\hsize}X
>{\hsize=.9\hsize\linewidth=\hsize}X
>{\hsize=.4\hsize\linewidth=\hsize}X
>{\hsize=.4\hsize\linewidth=\hsize}X}
    \hline\hline
          $(m_1,m_2)\,[M_\odot]$ & $(\beta_1,\beta_2)$ & $\eta$ & $M_B\,[M_\odot]$ & $\chi_1$ & $\chi_2$\\
    \hline
    $(6.9,4.8)$ & $(0.057,0.040)$ & $0.242$ & $120$ & $0.20$ & $0.10$\\
    $(6.4,5.2)$ & $(0.056,0.045)$ & $0.247$ & $115$ & $0.05$ &$0.35$\\
    $(13.8,9.6)$ & $(0.055,0.039)$ & $0.242$ & $250$ & $0.20$ & $0.10$\\
     \hline\hline
  \end{tabularx}
 
  \caption{ \normalsize Intrinsic parameters of the injected signals.
    }
    \label{tab:parameters}
\end{table}
    
The corner plots in Fig.~\ref{fig:corner1} show the reconstructed posterior distributions, together with the injected values, for the intrinsic parameters of the two lower-mass binaries in Table~\ref{tab:parameters}. The two-dimensional posteriors 
highlight a correlation between the symmetric mass ratio $\eta$ and the coupling $M_B$. This might be related to a correlation between 
$M_B$ and $\beta_2$ and to its anti-correlation with $\beta_1$.
The inferred values of the parameters (90\% confidence interval) 
are reported in Table~\ref{tab:errors}. As expected the chirp 
mass and the symmetric mass ratio are the best recovered 
parameters, 
but it is interesting that the BS coupling $M_B$ can be measured at the percent level, significantly better than the individual spins.

The results of the parameter estimation for the third system are shown in Fig.~\ref{fig:corner2}. This source has a larger chirp mass and a smaller Roche frequency ($f_{\rm Roche}\approx 45{\rm Hz}$), so the binary performs less cycles in the detector bandwidth. Correspondingly, we expect a worse recovery of the binary parameters. Indeed, the chirp mass is measured with larger uncertainty compared to the two systems with $\mathcal{M}=5M_\odot$, and error on the primary spin is roughly doubled compared to the first system. Interestingly, $M_B$ is still measured with very good accuracy, with relative errors below $3\%$. 
Overall, our results for the Bayesian inference are in good agreement with the estimates in~\cite{Pacilio:2020jza} based on a Fisher matrix analysis.

\begin{widetext}
\begin{center}
\begin{figure}[thp]
\centering
    \includegraphics[width=0.49\textwidth]{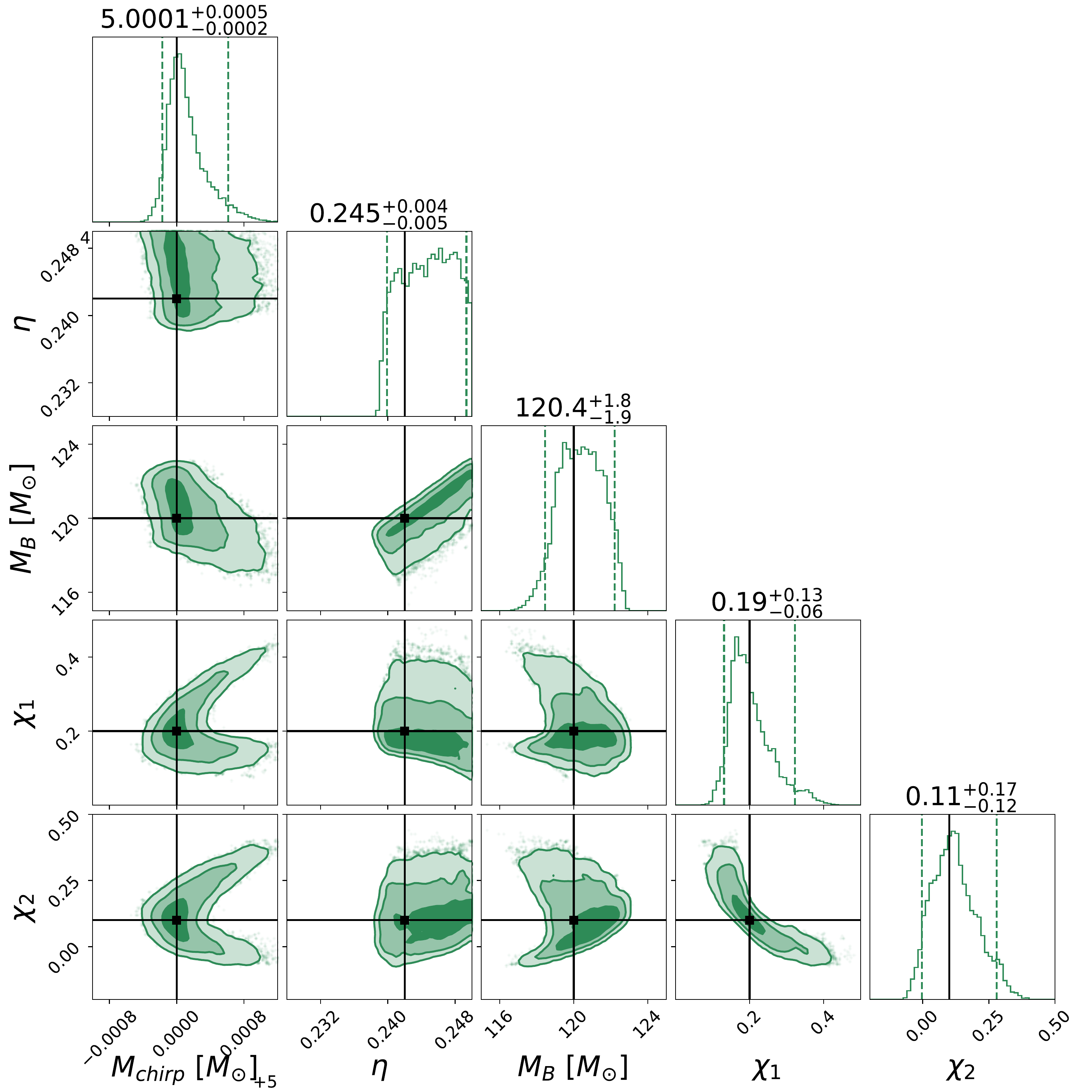}
    \includegraphics[width=0.49\textwidth]{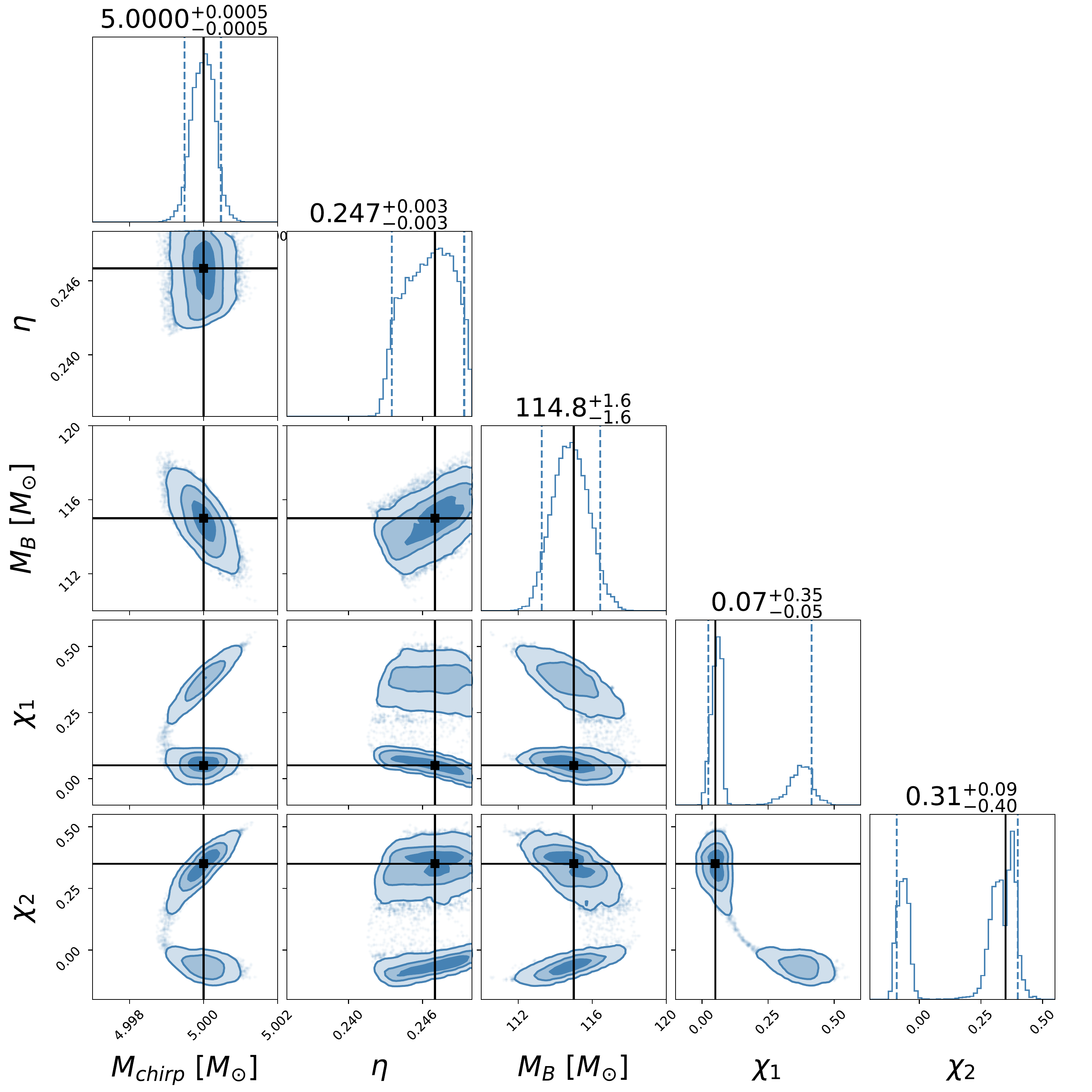}
    \caption{Contour plots for the recovered posterior distributions of two BS 
    binary signals corresponding to the 
    first two configurations shown in Table~\ref{tab:parameters}. Both systems have 
    injected chirp mass $\mathcal{M}=5M_\odot$, and 
    and ${\rm SNR}\sim130$. 
    (Left panel) The injected signal has $\eta=0.2423$ ($q\sim0.7$), $M_B=115M_\odot$ and $(\chi_1,\chi_2)=(0.2,0.1)$. 
    (Right panel) 
    The injected signal has $\eta=0.247$ ($q\sim0.8$), $M_B=115M_\odot$ and $(\chi_1,\chi_2)=0.05,0.35$. 
    A secondary peak, roughly in correspondence of the injected value for the other spin component, is evident in the posterior distributions for both $\chi_1$ and $\chi_2$ (see discussion below).}
    \label{fig:corner1}
\end{figure}
\end{center}
\begin{table}[thp]
\scriptsize
\begin{tabularx}{\linewidth}{
>{\hsize=1.6\hsize\linewidth=\hsize}X
>{\hsize=.7\hsize\linewidth=\hsize}X
>{\hsize=.9\hsize\linewidth=\hsize}X
>{\hsize=1.4\hsize\linewidth=\hsize}X
>{\hsize=1.4\hsize\linewidth=\hsize}X
>{\hsize=1.4\hsize\linewidth=\hsize}X
>{\hsize=.5\hsize\linewidth=\hsize}X
>{\hsize=1.\hsize\linewidth=\hsize}X
>{\hsize=.5\hsize\linewidth=\hsize}X
>{\hsize=1.\hsize\linewidth=\hsize}X
>{\hsize=.6\hsize\linewidth=\hsize}X}
    \hline\hline
          $(m_1,m_2)\,[M_\odot]$ & $d_L \,[{\rm Mpc}]$ & $f_{\rm Roche}\, [{\rm Hz}]$ & $\,\quad\mathcal{M}[M_\odot]$ & $\quad\quad\eta$ & $\,\quad M_B[M_\odot]$& $\delta {M_B}[\%]$ & $\,\quad\chi_1$ &$\delta {\chi_1}[\%]$ & $\,\quad\chi_2$ &$\delta {\chi_2}[\%]$\\
    \hline
    $(6.9,4.8)$ & $610$ & $100$ & $(4.9999,5.0006)$ & $(0.241,0.249)$ & $(118.5,122.2)$& $1.5\%$ & $(0.13,0.32)$& $50\%$ & $(-0.01,0.28)$&$131\%$\\
    $(6.4,5.2)$ & $600$ & $127$ & $(4.9995,5.0005)$ & $(0.244,0.250)$ & $(113.2,116.4)$& $1.5\%$ & $(0.02,0.42)$& $285\%$& $(-0.09,0.40)$ &$79\%$\\
    $(13.8,9.6)$ & $1150$ & $45$ & $(9.999,10.011)$ &$(0.237,0.249)$& $(241.5,253.9)$& $2.5\%$ & $(0.14,0.54)$& $83\%$ & $(-0.02,0.41)$& $134\%$\\
    
     \hline\hline
  \end{tabularx}
    \caption{\normalsize 90\% confidence intervals (CI) and relative percentage errors, expressed as the half-width of the CI with respect to the median of the distribution,
    for the recovered parameters of the three systems in Table~\ref{tab:parameters}.}
    \label{tab:errors}
\end{table}
\end{widetext}

\begin{figure}[thp]
\centering
    \includegraphics[width=0.49\textwidth]{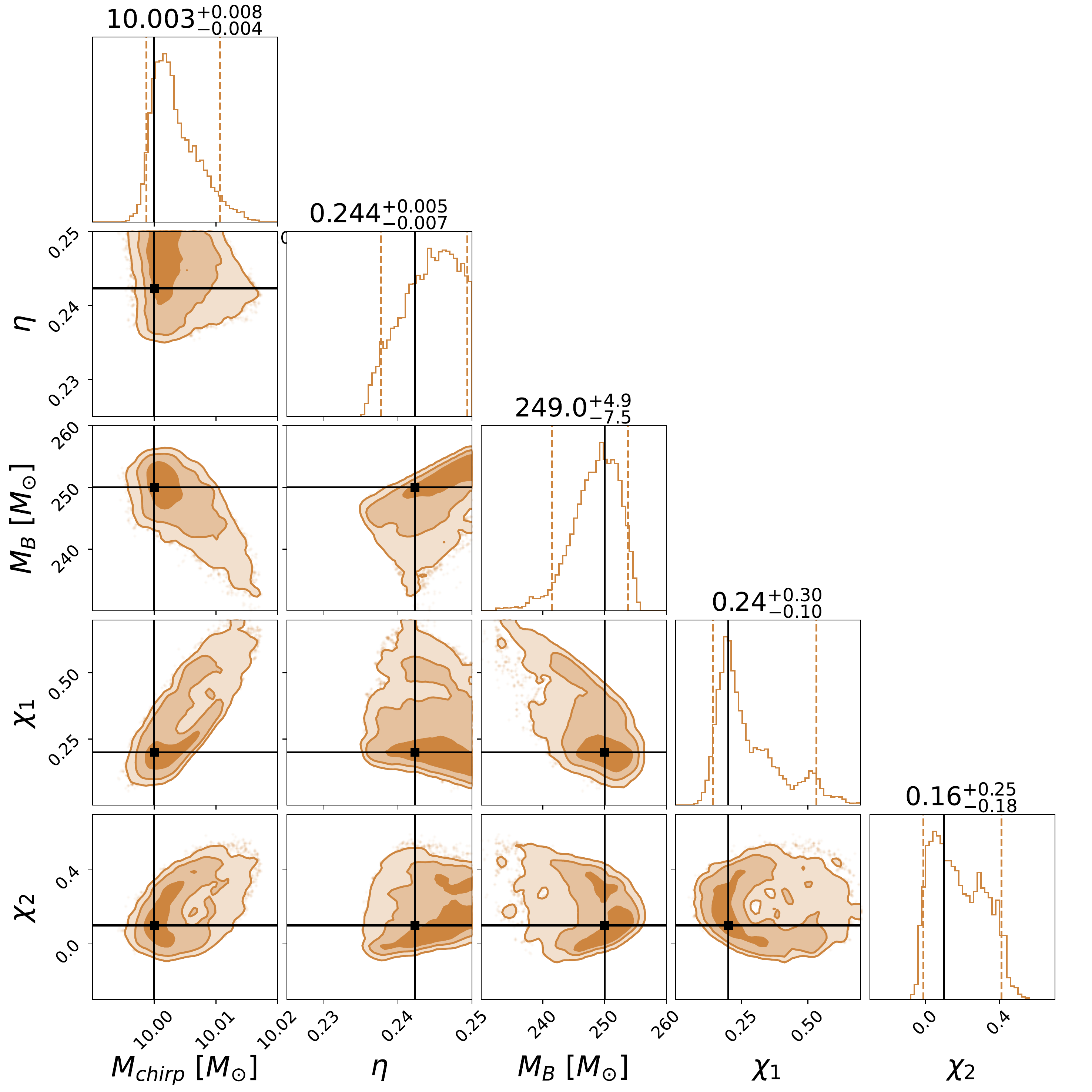}
    \caption{Same as Fig.~\ref{fig:corner1} but for 
    the third system in Table~\ref{tab:parameters}, 
    with a chirp mass $\mathcal{M}=10M_\odot$. 
    }
    \label{fig:corner2}
\end{figure}

A novel effect which emerges from the full Bayesian approach is that, for certain systems and choices of $\chi_1$ and $\chi_2$, there are degeneracies among different spin configurations and multipeak posteriors, which worsen the overall recovery. 
This effect appears to be more relevant for nearly symmetric 
binaries with moderate or high values of one of the spin components (or both)
for which the \emph{sign} of the antisymmetric spin combination, $\chi_a/|\chi_a|$,
is poorly measured. Indeed, in the equal mass case $\chi_a$ is completely 
undetermined as it only enters in the waveform through 
quadratic combinations, $\chi_a^2$, or in products 
$(\chi_1+\chi_2)(\kappa_2^1-\kappa_2^2)\chi_a$, 
both invariant under the exchange 
 $\chi_1\leftrightarrow \chi_2$.
This leads to two peaks in the posterior spin distributions, associated to the two possible signs $\chi_a=\pm|\chi_a|$. 

A representative example of these secondary peaks is contained in the right corner plot in Fig~\ref{fig:corner1}, which displays the marginalized posterior distribution of the individual spins for the second system in Table~\ref{tab:parameters}, having $\eta=0.247$ ($q=0.8$) and $\chi_1=0.05$, $\chi_2=0.35$. Although the injected parameters fall 
within the recovered posteriors, a small secondary peak in the distribution of 
$\chi_1$ contributes to a increase the 
error, while a higher one is also evident in the distribution of $\chi_2$. For fixed spin configurations, the bimodal feature is softened and tends to disappear for low mass ratios and high SNR.

\begin{figure}[thp]
\centering
    \includegraphics[width=0.49\textwidth]{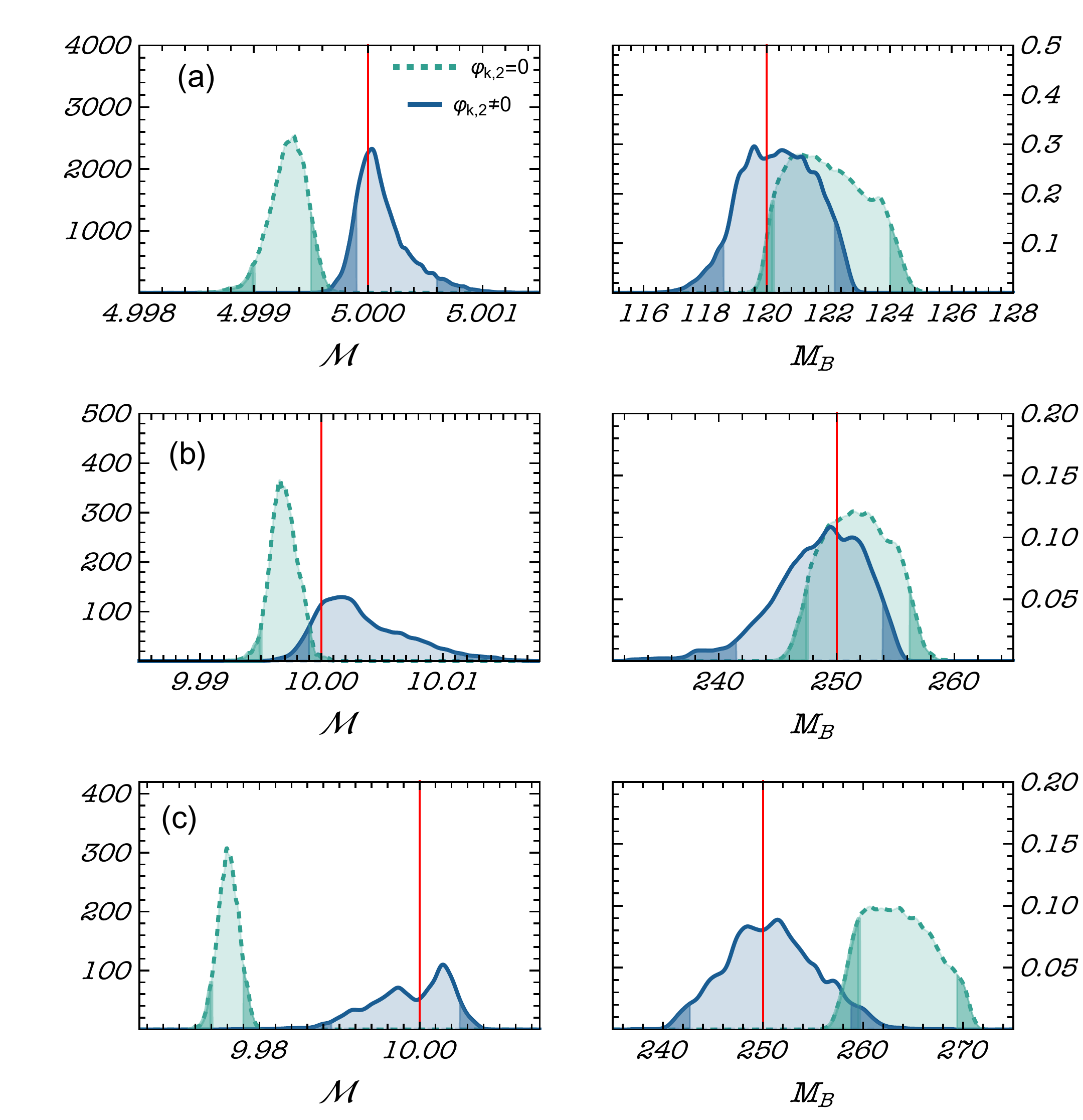}
    \caption{Posterior distributions of $\mathcal{M}$ (left panels) and $M_B$ (right panels) for different systems, obtained 
    fixing $\varphi_{2,\kappa}=0$ in the recovery (dashed) or using the full template (solid). The 
    injected values are identified by a vertical red line while CI of $90\%$ are marked by a change in the curve filling. The considered systems are: (a) the first system in Table~\ref{tab:parameters}, (b) the third system in Table~\ref{tab:parameters}, (c) the same system as in (b) but with intermediate spins $(\chi_1,\chi_2)=(0.4,0.7)$}
    \label{fig:noquadrupole}
\end{figure}

Excluding situations where such degeneracy is present, we found that including quadrupolar corrections generically 
improves the parameter estimation. Indeed, although the constraints on $M_B$ come mostly from the tidal deformability contribution, overall the errors on all parameters increase when setting the quadrupole corrections to zero. The reconstruction of $\chi_1$ and $\chi_2$ is, as expected, the most affected and their uncertainty typically increases by a factor of $\sim 4$ when only including point-particle terms up to $3.5$pN and tidal corrections. Furthermore we point out that neglecting the quadrupole moment in the recovery leads to biases in the measurement of the parameters if the signal is injected using the full waveform template. Figure~\ref{fig:noquadrupole} shows this case for three different binaries, respectively corresponding to: the first and third systems in Table~\ref{tab:parameters} and a system having $\mathcal{M}=10M_\odot$ (i.e. with the same chirp mass of our third system) but intermediate spins, $(\chi_1,\chi_2)=(0.4,0.7)$. 
The estimated value of the chirp mass is always biased, the bias 
being more relevant for the low-mass (panel a) and intermediate-spin (panel c) systems. In the first case this is due to the overall 
better accuracy in the recovery. In the case of system (c) 
the higher values of the spins imply larger quadrupolar 
corrections (consistent with the bottom panels of 
Fig.~\ref{fig:pn_contribution}) and thus a larger departure from the original 
signal when neglecting them. The bias on $M_B$ is weaker 
but still relevant for systems (a) and (c), while the 
inferred values of $M_B$ for system (b) are fully compatible in the two 
cases. 

\section{Discussion and conclusions}
We have presented the first Bayesian inference on BS binaries with large self interactions using a coherent inspiral template. Compared to the standard binary BH case, this template contains an extra parameter that regulates both the tidal deformability and the spin-induced quadrupole moment in terms of the masses and spins of the binary components.

Making a conservative assumption and cutting the signal at the threshold for tidal disruption, we found that future third-generation detectors such as ET can measure the BS coupling constant at the percent level. 
The  best constraints come from binaries with relatively small chirp mass, which are also those for which the inspiral-only template is more relevant.

Overall, we confirm the estimates in Ref.~\cite{Pacilio:2020jza} obtained by a Fisher-matrix analysis and with an approximated spin-induced quadrupole moment. Using the corroborated estimates of Ref.~\cite{Pacilio:2020jza} as a guide, we expect the same percent accuracy in measuring $M_B$ also with the future space mission LISA~\cite{Barausse:2020rsu,LISA:2022kgy} but for BSs with masses around $10^5 M_\odot$ (corresponding to $M_B\approx(10^5-10^6)M_\odot$) at a luminosity distance of $10\,{\rm Gpc}$.

Given the relatively small cut-off frequency, the accuracy of this measurement relies on ET sensitivity around and below $100\,{\rm Hz}$ and it would quickly deteriorate for current detectors. Nevertheless, it might be interesting to apply our template to real LIGO-Virgo-KAGRA data, also comparing the Bayesian evidence for the BS model with that for a standard BH binary inspiral.
Along this line --~and since our template is based on the TaylorF2 approximant and therefore valid only in the inspiral phase~-- natural extensions of our work include using this template to inform effective inspiral-merger-ringdown models~\cite{Toubiana:2020lzd} or to extend merger waveforms obtained from numerical-relativity simulations~\cite{Liebling:2012fv,Palenzuela:2007dm,Palenzuela:2017kcg,Bezares:2017mzk,Sanchis-Gual:2018oui,Bezares:2022obu,Siemonsen:2023hko} (as in the recent case of Proca star head-on collisions~\cite{Sanchis-Gual:2022mkk} and the subsequent Bayesian inference~\cite{2022arXiv220602551C}).
Another possibility would be extending the waveform using effective models with frequency-dependent tidal deformability terms, as recently done in Ref.~\cite{DeLuca:2022xlz}.

\begin{acknowledgments}
Numerical calculations have been performed at the CINECA Marconi cluster through a CINECA-INFN agreement.
P.P. acknowledge financial support provided under the European
Union's H2020 ERC, Starting Grant agreement no.~DarkGRA--757480 and under the MIUR PRIN programme, and support from the Amaldi Research Center funded by the MIUR program ``Dipartimento di Eccellenza" (CUP:~B81I18001170001). 
C.P.~is supported by European Union's H2020 ERC Starting Grant No. 945155--GWmining and by Cariplo Foundation Grant No. 2021-0555.
\end{acknowledgments}

\appendix
 
\section{Waveform phase coefficients}\label{sec:pnappendix}
Here we show the explicit expressions of the pN coefficients of the waveform phase expansion of Eq.~\eqref{eq:PN_expansion} in terms of the masses $m_i$, the dimensionless spins $\chi_i=J_i/M_i^2$, the tidal deformabilities $\Lambda_i=(2/3)\lambda_2^i\mathcal{C}_i^{-5}$ and the reduced quadrupole moments $\kappa_2^i=Q_i/(\chi_i^2m_i^3)$. \\
\\
The non-zero coefficients are listed below, where the following quantities have been introduced
\begin{align*}
&m=m_1+m_2 \qquad &&\eta=m_1m_2/M^2\\  &\delta=(m_1-m_2)/M \qquad &&\chi_{s/a}=(\chi_1\pm\chi_2)/2\\  &\Lambda_{s/a}=(\Lambda_1\pm\Lambda_2)/2 \qquad && \kappa_{s/a}=(\kappa_2^1\pm\kappa_2^2)/2\\
&\gamma_E=0.5772156649. 
\end{align*}
We considered spin-spin and spin-orbit corrections up to second order in the spin. 
\begin{widetext}
\[
\begin{aligned}
\varphi_{{\rm pp},0}&=1\\
\varphi_{{\rm pp},0.5}&=0\\
\varphi_{{\rm pp},1}&=\frac{3715}{756}+\frac{55\eta}{9}\\
\varphi_{{\rm pp},1.5}&=-16\pi+\frac{113}{3}\delta \chi_a+\Bigl(\frac{113}{3}-\frac{76\eta}{3}\Bigr)\chi_s\\
\varphi_{{\rm pp},2} &= \frac{15293365}{508032}+\frac{27145 \eta}{504}+\frac{3085 \eta^{2}}{72}-\frac{5}{8}\chi_{\mathrm{s}}^{2}\left(1+156 \eta\right)
+\chi_{\mathrm{a}}^{2}\left(-\frac{5}{8}+100 \eta \right)-\frac{5}{4}\chi_{\mathrm{a}}\chi_{\mathrm{s}}\delta, \\
\varphi_{{\rm pp},2.5}&= [1+\log (\pi M f)]\left[\frac{38645 \pi}{756}-\frac{65 \pi \eta}{9}+\delta\left(-\frac{732985}{2268}-\frac{140 \eta}{9}\right) \chi_{a}+\left(-\frac{732985}{2268}+\frac{24260 \eta}{81}+\frac{340 \eta^{2}}{9}\right) \chi_{s}\right]\\
\varphi_{{\rm pp},3}&= \frac{11583231236531}{4694215680}-\frac{6848 \gamma_{E}}{21}-\frac{640 \pi^{2}}{3}+\left(-\frac{15737765635}{3048192}+\frac{2255 \pi^{2}}{12}\right) \eta+\frac{76055 \eta^{2}}{1728}-\frac{127825 \eta^{3}}{1296} \\
&-\frac{6848}{63} \log (64 \pi M f)+\pi\left[ \frac{2270}{3} \delta \chi_{\mathrm{a}}+\left(\frac{2270}{3}-520 \eta\right)\chi_{\mathrm{s}} \right]\\
&+\chi_{\mathrm{s}}^{2}\left(-\frac{1344475}{2016}+\frac{829705}{504} \eta+\frac{3415}{9} \eta^{2}\right)+\chi_{\mathrm{a}}^{2}\left(-\frac{1344475}{2016}+\frac{267815}{252} \eta-240 \eta^{2}\right)+\chi_{\mathrm{a}}\chi_{\mathrm{s}}\delta\left(-\frac{1344475}{1008}+\frac{745}{18} \eta\right)\\
\varphi_{{\rm pp},3.5} &= \frac{77096675 \pi}{254016}+\frac{378515 \pi \eta}{1512}-\frac{74045 \pi \eta^{2}}{756}+\left(-\frac{25150083775}{3048192}+\frac{26804935}{6048} \eta-\frac{1985}{48} \eta^{2}\right) \delta \chi_{\mathrm{a}}\\
&+\left(-\frac{25150083775}{3048192}+\frac{10566655595}{762048} \eta-\frac{1042165}{3024} \eta^{2}+\frac{5345}{36} \eta^{3}\right) \chi_{\mathrm{s}}\\
\varphi_{\kappa,2} &= -\frac{5}{8}\chi_{\mathrm{s}}^{2}\left[80 \delta \kappa_{a}+80(1-2 \eta) \kappa_{s}\right] +\chi_{\mathrm{a}}^{2}\left[-50 \delta \kappa_{a}-50 \kappa_{s}+100 \eta\kappa_{s}\right]-\frac{5}{4}\chi_{\mathrm{a}}\chi_{\mathrm{s}}\left[80(1-2 \eta) \kappa_{a}+80 \delta \kappa_{s}\right], \\
\varphi_{\kappa,3}&=\left(\chi_{\mathrm{s}}^{2}+\chi_{\mathrm{a}}^{2}\right)\left[\delta\left(\frac{26015}{28}-\frac{1495}{6} \eta\right) \kappa_{a}+\left(\frac{26015}{28}-\frac{44255}{21} \eta-240 \eta^{2}\right) \kappa_{s}\right]\\
&+\chi_{\mathrm{a}}\chi_{\mathrm{s}}\left[\left(\frac{26015}{14}-\frac{88510}{21} \eta-480 \eta^{2}\right) \kappa_{a} +\delta\left[\left(\frac{26015}{14}-\frac{1495}{3} \eta\right) \kappa_{s}\right]\right]\\
\varphi_{\tn{T},5}&=-24[(1+7\eta-31\eta^2)\Lambda_s+\delta(1+9\eta-11\eta^2)\Lambda_a], \\
\varphi_{\tn{T},6}&=-\frac{3115}{52}[(1+7\eta-31\eta^2)\Lambda_s+\delta(1+9\eta-11\eta^2)\Lambda_a]\\
&+\frac{6595}{364}\delta\Bigl[\delta\Bigl(1-\frac{13272}{1319}\eta+\frac{8944}{1319}\eta^2\Bigr)\Lambda_s+\Bigl(1-\frac{15910}{1319}\eta+\frac{32850}{1319}\eta^2+\frac{3380}{1319}\eta^3\Bigr)\Lambda_a \Bigr]\,.\\
\end{aligned}
\]
\end{widetext}

\section{On the $\Lambda(\beta)$ and $\kappa_2(\chi,\beta)$ relations}\label{sec:fitappendix}
To obtain the tidal deformability we inverted the relation for $\beta(\Lambda)$ provided in Ref.~\cite{Sennett:2017etc}, for which the authors estimate an error of the order of $1\%$. However, the error on the inverse can be higher, especially close to the stationary point at $\beta\approx0.06011$, where the first derivative of the inverse function formally diverges. Neglecting the error introduced in numerically evaluating the inverse, we can simply consider the propagation of the relative error on $\beta$:
\begin{equation*}
\delta_{\rm rel}\Lambda = \frac{\beta\cdot\Lambda'(\beta)}{\Lambda(\beta)} \delta_{\rm rel}\beta
\end{equation*}
 Applying this relation, the error on $\Lambda$ is below $10\%$ for $\beta \leq 0.057$, which is saturated by the higher $\beta$ considered for the primary object in the main text. The stationary point is excluded even from the sampling because of the constraint $\beta\leq 0.06$. 
 
To obtain the reduced quadrupole moment $\kappa_2(\chi,\beta)$ we interpolated the data in~\cite{Vaglio:2022flq} on a bi-dimensional grid $n_\beta\times n_\chi=9\times30$, using cubic splines. An error bound for cubic spline interpolation is provided in Ref.~\cite{hall_optimal_1976}
\begin{equation*}
\delta s = |f-s| \sim \frac{3}{154}|f^{(4)}|h^4 \sim  \frac{3}{154}\left|\frac{s''_{i+1}-2s''_i+s''_{i-1}}{h^2}\right| h^4  
\end{equation*}
where $s$ is the spline interpolation function, $s''_i$ its second derivative at a given grid point and $h$ the grid spacing. We used equally-spaced grid points in $\beta$ and in the logarithm of the spin, $\log(\chi)$. We can apply the formula separately for the interpolation in $\beta$ and $\chi$, using $h_\beta=\beta_{i+1}-\beta_{i}$ and $h_\chi=\log(\chi_{i+1})-\log(\chi_i)$. 
We found that the errors are of the order of $1\%$ and $5\%$ in the $\beta$ and $\chi$ direction, respectively. For spin larger than $\chi\sim 0.1$, this will be the dominant contribution to the total uncertainty on the reduced quadrupole $\kappa_2$, while for smaller $\chi$, it will be dominated by the numerical error on the data, which can reach the order of $10\%$ for such small spin values~\cite{Vaglio:2022flq}. 

\bibliography{refs}

\end{document}